\documentstyle[aps,preprint,epsf]{revtex}
\tightenlines
\begin{document}

%
%
\title{Remnants of Initial Anisotropic High Energy Density Domains
	in Nucleus-Nucleus Collisions}
\author{F.~Wang$^{a}$ and H.~Sorge$^{b}$}
\address{\em (a) Nuclear Science Division, LBNL, Berkeley, CA 94720, USA}
\address{\em (b) Department of Physics, SUNY at Stony Brook, NY 11794, USA}
\maketitle

%
%
\bigskip
\begin{abstract}
Anisotropic high energy density domains may be formed at early stages of
ultrarelativistic heavy ion collisions, e.g. due to phase transition
dynamics or non-equilibrium phenomena like (mini-)jets. Here we  investigate
hadronic observables resulting from an initially created anisotropic high 
energy density domain. Based on our studies using a transport model we find 
that the initial anisotropies are reflected in the freeze-out multiplicity 
distribution of both pions and kaons due to secondary hadronic rescattering.
The anisotropy appears to be stronger for particles at high transverse
momenta. The overall kaon multiplicity increases with large fluctuations
of local energy densities, while no change has been found in the
pion multiplicity.  
\end{abstract}


%
%
Under normal conditions, quarks and gluons are confined in hadrons.
The situation is entirely different 
at  high  densities as may have been created  in the early universe
shortly  after the Big Bang  or in today's laboratory experiments
in which two nuclei are colliding. It is impossible under such conditions
that spatially separated
hadrons be formed. Instead, formation of quark matter is 
expected~\cite{Lee74:qgp,Kei76:star,Raf82:qgp,McL86:qgp,qm91:Mul}.
The primary goal of high energy heavy ion experiments is to create
this new state of matter, the quark-gluon-plasma (QGP) in which quarks
and gluons are no longer confined in individual hadrons.  
It is conceivable that a deconfined matter
may be formed at an early stage of a heavy ion collision
only in a localized region in configuration space,
surrounded by hadronic matter.
For example, Gyulassy {\em et~al.}~\cite{Gyu97:hot-spots} argued
that as a result of multiple mini-jet production at RHIC energy,
the initial conditions of the QGP formed in ultrarelativistic
nuclear collisions might be inhomogeneous, with large fluctuations
of the local energy density.
They showed that such initial conditions could result in azimuthal
asymmetry in particle distributions.
In contrast to mini-jet production,
S.~Mr\'{o}wczy\'{n}ski~\cite{Mro97:plb} argued that color filamentation
could also result in azimuthal asymmetry by a flow of large number of
particles with relatively small transverse momenta.
Kapusta {\em et~al.}~\cite{Kap95:nucleation}
argued that the probability of a hard nucleon-nucleon
collision that was able to nucleate a seed of QGP in the surrounding
hot and dense hadronic matter in a central heavy ion collision may not
be small even at AGS energy.
In addition, disoriented chiral condensates (DCC) may be formed in heavy
ion collisions by the restoration of chiral
symmetry~\cite{Raj93:npb399,Raj93:npb404}.
The formation of DCC could result in a spectacular asymmetry in charged
(and neutral) pion phase-space distributions~\cite{Gav94:prl}.

Once created at early stage of a heavy ion collision,
a QGP will undergo a phase transition from the deconfined state to
a normal hadonic state, followed by hadronic expansion.
What are measured in experiments are the particle momenta  long after
these particles have ceased to interact (freeze-out).
It remains an experimental and theoretical challenge to extract
information on the initial conditions 
from the particle distributions at freeze-out.
Early investigation by Pratt~\cite{Pra94:droplet} showed that two
particle correlations may be useful to search for these fluctuations
in initial conditions.

Because of secondary particle interactions (hadronic rescattering),
primordial quark-gluon distributions from a QGP may be completely altered
during the hadronic expansion stage.
On the other hand, an initial anisotropy
from the quark-gluon stage in configuration space  may 
not be washed out but survive
the hadronic stages. 
Non-isotropic energy densities may leave
their imprint on  momentum spectra due to hadronic rescattering. 
Different amounts of rescattering 
can result in variations of  particle multiplicities along 
different   directions.
An interesting consequence may be that a non-spherical
high energy density domain, which can be formed at an early stage
of a heavy ion collision due to fluctuations in energy deposition,
may produce an anisotropy in pion and kaon emission after freeze-out.
It is the goal of the present work to study the possible
magnitude of such an anisotropy, given the magnitude of the
initial anisotropy in configuration space.

Since we are going to study 
the translation of initial spatial anisotropies into final-state
momentum  anisotropies  of kaons and pions
due to  hadronic rescattering, it is important
that the model to be used contains the relevant dynamics and describes
existing experimental data to a reasonable degree.
One of the important physics results learned from current heavy ion 
experiments at the BNL AGS and the CERN SPS
is that there is a large amount of secondary particle interactions
in central collisions.
Transport models~\cite{qm88:Sor:rqmd,Sor89:rqmd,Sor95:flavor-changing,Win96:QM96,Kah92:strangeness,Li95:dense}
have been employed to study the importance of
these  interactions for observables.                
For example,
strangeness yields may change drastically due to the hadronic 
cascading~\cite{Mat89:rqmd-k-pi,Sor92:rqmd-ropes,Ber94:PLB}.
In fact, the systematics of the observed strangeness 
enhancement~\cite{Abb90:e802-kaon-pion,Abb92:e802-cent,Bar90:na35_zpc,Bor97:sqm97}
-- energy and mass dependence --
can be described by the Relativistic Quantum Molecular Dynamics (RQMD)
approach~\cite{Sor91:rqmd-meson,Sor95:ZfP,Xu98:omega} rather well.
In this paper, we use the RQMD approach to investigate
the translation of initial spatial anisotropies into final-state
momentum  anisotropies  of kaons and pions.
One should note, however, that the excess of strange anti-baryons
 which has been 
observed~\cite{Bar90:na35_zpc,Sto91:npa,Aba91:wa85_plb}
cannot be solely produced by hadronic
collisions within the RQMD approach~\cite{Sor94:qm93}.
More complicated mechanisms pointing to earlier and denser
stages are required. For instance, formation of
color ropes~\cite{Sor92:rqmd-ropes,Sor94:qm93,Sor95:qm95}
which has been implemented in RQMD, 
density-dependent antibaryon masses~\cite{Ko92:prc} or quark
coalescence~\cite{Csi98:alcor,Zim97:alcor}
may be responsible.
While the strange antibaryons may well be a very interesting probe
of the matter deposition at early times in heavy ion collisions,
it should be noted that they are very rare particles at presently
accessible energies~\cite{talk:Koc90}.
Lack of statistics has prevented us to employ these species
in the context discussed here
of  dynamically caused  event-by-event fluctuations.

RQMD (version 2.3)~\cite{Sor95:flavor-changing} 
was used to simulate head-on Pb+Pb
collisions (impact parameter $b=0$) at SPS energy (158 GeV/nucleon).
The influence of the hadronic rescattering on the inclusive yields
has already been studied by one of the authors. 
By including rescattering in RQMD, pion multiplicity 
is reduced by 8\% and kaon multiplicity increases by over 50\%
at SPS energy~\cite{Sor95:flavor-changing}. 
  
For the present work we did not attempt to include dynamical
fluctuations in the local energy deposition
other than given by the standard RQMD procedure. Instead, 
the initial conditions after hadronization as resulting from the model were
modified  to create artificially 
 anisotropic  domains beyond the model's intrinsic
-- dynamical and statistical -- fluctuations.  
We feel this is justified in view of 
the considerable uncertainties in the early-state dynamics which
cannot be reliably calculated based on quantum chromodynamics (QCD).
 
In the following we specify our strategy to include additional
fluctuations on an event-by-event basis: 
At $t=1$~fm/c in the center of mass frame,
a  randomly chosen fraction of hadrons with rapidity $|y|<1.5$
was moved into an elliptical area in the transverse plane,   
located around the center of the fireball,
and the longitudinal position ($z$) was kept unchanged.
The new $(x,y)$ positions in the cylinder were randomly sampled
from a homogeneous distribution.
Size of the transverse cross-section of the cylinder was chosen as
8~fm (major axis) $\times$ 2~fm (minor axis). 
The total energy and momentum are conserved by this procedure.
On the other side, the artificially induced fluctuations
of the local hadron densities modify 
the subsequent collision dynamics.
After the re-arrangement, RQMD is properly re-initialized
and propagates the hadrons until freeze-out.

We have studied two cases:
\begin{enumerate}
\item 10\% of the hadrons ($|y|<1.5$) were moved.
Approximately 90 hadrons  representing a total energy of 120 GeV
were moved in each event.
This resulted in a 30\% increase of total energy within the cylinder,
and a 20--30\% increase in the total number of binary collisions.
About 11K events were studied for this case.
\item 20\% of the  hadrons  ($|y|<1.5$) were moved.
Approximately 180  hadrons  representing a  total energy of 240 GeV
were moved in each event.
This resulted in an 60\% increase of total energy within the cylinder,
and a 40--50\% increase in the total number of binary collisions.
About 10K events were studied for this case.
\end{enumerate}

The left panel of Fig.~\ref{fig:conf} shows the 
projection of the particle density  onto the $z=0$ plane, 
resulting from the re-arrangement at $t=1$ fm in the center of mass frame.
The right panel of Fig.~\ref{fig:conf} shows the level of asymmetry
in configuration space due to the creation of a high energy density domain.
Fig.~\ref{fig:e_vs_t} shows average energy density
within the elliptical cylinder as a function of time in the center of mass
frame for the second case.
The solid curves are for energy summed over all particles/strings,
and the dashed ones are those for particles/strings within $|y|<1.5$. 
Artificial creation of the high energy density domain is visible as
 the discontinuity at $t=1$ fm.
The bases are the time profile of the average energy density from
a default RQMD event.
Although the energy density in the central elliptical cylinder
is artificially increased at the initial stage, 
the freeze-out density is similar to that from a default RQMD event,
as shown in the figure.
This indicates that the freeze-out density is constant in RQMD over
a wide range of initial conditions.

As discussed at the beginning, the purpose of the study is to
investigate possible signatures of high energy density domains at
initial stages of heavy ion collisions in the azimuthal distributions
of particles at freeze-out.
The resulting freeze-out charged pion distribution is displayed in
Fig.~\ref{fig:pion}, for the first case in the upper panel and for
the second case in the bottom panel.
As clearly seen from the figure, the distributions are
azimuthally asymmetric.
There are fewer pions emitted along the major axis of the ellipse,
in agreement with the picture that pion yield is reduced by
rescattering~\cite{Sor95:flavor-changing}. 
It should be noted that this result qualitatively agrees with 
the result that the elliptic flow is in-plane at SPS energy 
for medium impact parameters~\cite{App98:na49_prl_flow}
where the initial geometry is 
similar to the high energy density domain studied here.
Only particles with rapidities $|y|<2$ are plotted in the figure.
No anisotropy was found for particles with rapidity of more than 2 units
away from mid-rapidity.
This is presumably due to that only  particles
within $|y|<1.5$ were modified.

The azimuthal multiplicity distributions can be described
by the functional form
\begin{equation}
\frac{dN}{d\phi} \propto 1-\alpha\cos(2\phi). \label{eq}
\end{equation}
A fit to the distribution yields $\alpha=0.35\%$ and 0.69\%
for case 1 and case 2, respectively.
The magnitude of the anisotropy, $\alpha$, depends upon pion
transverse momentum, $p_{\perp}$ (GeV/c);
the higher $p_{\perp}$, the greater the anisotropy.
The azimuthal distributions for pions of different $p_{\perp}$
ranges are shown in Fig.~\ref{fig:pion_pt} for the second case.
The fit results of the anisotropy in pion multiplicity distribution
for various $p_{\perp}$ cuts, as well as the one including all $p_{\perp}$
in Fig~\ref{fig:pion},
are summarized in Table~\ref{table1}.
There is no difference observed between $\pi^+$ and $\pi^-$.
The $p_{\perp}$ dependence can be qualitatively understood. High $p_{\perp}$
pions colliding with other particles have high probability to be
destroyed to produce other particles, 
for instance, $\pi\pi\rightarrow KK$.
Low $p_{\perp}$ pions, on the other hand, more often
elastically scatter from other particles resulting in no change of
the pion multiplicity. 
The anisotropy must asymptotically approach zero  for
zero $p_{\perp}$ particles as required from symmetry considerations.

The azimuthal distribution of charged pion multiplicity ($|y|<2$)
from default RQMD events is constant as expected. 
The fitted average of the distribution from the default events
is shown as the dotted line on the plots in Fig.~\ref{fig:pion}
and Fig.~\ref{fig:pion_pt}.
There is no essential change in the overall charged pion multiplicity
due to the initial high energy density domain.
See Table~\ref{table2} where the relative changes are tabulated.

The azimuthal distributions of $K^+$ (left panel) and $K^-$
(right panel) multiplicities
are shown in Fig.~\ref{fig:kaon} for the second case.
A cut of $|y|<2$ is also applied in all plots.
The top two plots are the distributions for kaons including all $p_{\perp}$,
whereas the other plots have various $p_{\perp}$ cuts.
As can be seen from the plots, the kaon multiplicity is anisotropic
as well. The fits to the functional form~(\ref{eq}) are superimposed on
the figure as the solid curves. The fit results of the anisotropy 
are tabulated in Table~\ref{table1}.
The magnitude of the anisotropy in the kaon multiplicity is similar to
the pions. This indicates that absorption of pions 
and kaons is similar in the fireball.

However, compared to the kaon multiplicity from the default RQMD events,
the overall kaon multiplicity is increased.
The fitted average to the kaon multiplicity ($|y|<2$)
from default RQMD events
is shown as the straight line on each plot in Fig.~\ref{fig:kaon}.
The relative increase in $K^-$ is larger than in $K^+$,
whereas the absolute increase is similar. For reference,
the magnitudes of the absolute and relative increases in 
kaon multiplicity are tabulated in Tables~\ref{table2} and~\ref{table3},
respectively.
To demonstrate this point more clearly, the azimuthal distributions of
the difference in $K^+$ and $K^-$ multiplicities are computed and
plotted in the top panel of Fig.~\ref{fig:lambda}.
The solid curves are fits to the functional form~(\ref{eq}), and
the dotted lines are the fitted average to the corresponding distributions
from default RQMD events. As can be seen from the plots,
the increase in this quantity from default RQMD events is minor.
Presumably, the slight increase in this quantity from default RQMD events
is due to the excess in $K^+$ production (over $K^-$)
through associate production together with $\Lambda$'s.
To illustrate this point, azimuthal distributions of $\Lambda$
multiplicity are plotted in the bottom panel of Fig.~\ref{fig:lambda}.
The distributions of $N_{K^+} - N_{K^-}$ and $\Lambda$ are indeed similar.

The following picture emerges from the above results.
More pions are absorbed along the major axis of the elliptical
high energy density domain.
Kaons are absorbed in the same manner as pions.
The increase in number of binary collisions, presumably, enhances
particle production including both pions and kaons.
Large fraction of these pions, however, are destroyed by rescattering
producing additional kaons.
These kaons are produced dominantly through pair production mechanism.
There are slight excess in $K^+$ over $K^-$ from associate production
of $K^+$ together with $\Lambda$'s, mainly in the high $p_{\perp}$ region.

In summary, we have quantitatively studied, 
in  the framework of RQMD model, 
the remnants of non-spherical, high energy
density domains in particle emissions at freeze-out. 
Such domains can produce anisotropy in charged pion and kaon
emission at freeze-out due to particle rescattering;
the anisotropy persists more in high $p_{\perp}$ particles.
For a 27\% asymmetry in the initial configuration distribution
(for the second case), a 0.7\% anisotropy was found in the pion
distribution at freeze-out.
This anisotropy increased from 0.4\% for pions with $p_{\perp}<0.5$ GeV/c
to 2.2\% for pions with $p_{\perp}>0.8$ GeV/c.
The  factor how the initial asymmetry
in configuration space translates into final momentum space appears to be
rather small. Depending on the transverse momentum window it
varies between 1.5 and 8 percent.
The anisotropy shows similar strength in kaon distributions.
The addition of the high energy density domain resulted in enhanced
kaon multiplicity at freeze-out, while there was essentially no change
in the overall pion multiplicity.
The increase in kaon multiplicity is mainly due to increased
pair production.

The current study used RQMD events simulated with exact zero impact
parameter to avoid possible anisotropy in the freeze-out particle
multiplicity distribution arising from finite impact parameter.
The magnitude of such anisotropy in central collisions at small
finite impact parameters has not been studied in details.
The current study assumed that the high energy density domain was present
in every event at the center and had the same size and shape.
It remains a future task to study the dependence of the anisotropy 
as functions of location, size and shape of the high energy density domain.
In the present work, the anisotropy in particle emission
is studied only inclusively.
It remains a challenge to study the effect on an event-by-event basis,
given the relative weakness of the effect as shown by the present work.

%
One of the authors (FW) thanks Dr.~P.~Jacobs, Dr.~A.~M.~Poskanzer,
Dr.~H.~G.~Ritter and Dr.~N.~Xu for valuable discussions.
He also thanks the Institute for Nuclear Theory
where part of this work was carried out.
This work was supported by the U.~S.~Department of Energy
under contracts DE-AC03-76SF00098 and DE-FG02-88ER40388.
This research used resources of the National Energy Research
Scientific Computing Center.

%
%
\bibliography{paper_ref}
\bibliographystyle{phunsrt}

%
%

%
%
\pagebreak

\begin{table}
\caption{Pion and kaon emission anisotropy at freeze-out, $\alpha$,
resulting from the initial anisotropic high energy density domain in 
central Pb+Pb collisions (impact parameter $b=0$ fm) at 158 GeV/nucleon
simulated by RQMD.}
\label{table1}
\begin{tabular}{c|c|r|r|r|r} \hline
       & particle & all $p_{\perp}$ & $p_{\perp}<0.5$ 
       & $p_{\perp}>0.5$ & $p_{\perp}>0.8$ \\ \hline
       & $\pi$ & 0.4\% & 0.2\% & 0.5\% & 1.0\% \\
case 1 & $K^+$ & 0.3\% & 0.0\% & 0.4\% & 0.7\% \\
       & $K^-$ & 0.4\% & 0.0\% & 0.8\% & 0.7\% \\ \hline
       & $\pi$ & 0.7\% & 0.4\% & 1.1\% & 2.2\% \\
case 2 & $K^+$ & 1.1\% & 0.6\% & 1.2\% & 1.9\% \\
       & $K^-$ & 1.0\% & 0.3\% & 1.3\% & 2.0\% \\ \hline
\end{tabular}
\end{table}

\begin{table}
\caption{Relative increase in pion and kaon multiplicity due to
the initial anisotropic high energy density domain from default RQMD events.}
\label{table2}
\begin{tabular}{c|c|r|r|r|r} \hline
       & particle & all $p_{\perp}$ & $p_{\perp}<0.5$ 
       & $p_{\perp}>0.5$ & $p_{\perp}>0.8$ \\ \hline
       & $\pi$ & $-0.3$\% & $-0.3$\% & $-0.1$\% & $-0.1$\% \\
case 1 & $K^+$ &  4.7\% &  3.4\% &  4.7\% &  8.0\% \\
       & $K^-$ &  6.7\% &  3.9\% &  8.2\% &   11\% \\ \hline
       & $\pi$ & $-0.7$\% & $-1.2$\% & 0.2\% & 1.9\% \\
case 2 & $K^+$ &  3.8\% &  1.8\% & 4.0\% & 9.0\% \\
       & $K^-$ &  5.4\% &  1.6\% & 6.8\% &  12\% \\ \hline
\end{tabular}
\end{table}

\begin{table}
\caption{Absolute increase in kaon multiplicity due to
the initial anisotropic high energy density domain from default RQMD events.}
\label{table3}
\begin{tabular}{c|c|r|r|r|r} \hline
       & particle & all $p_{\perp}$ & $p_{\perp}<0.5$ 
       & $p_{\perp}>0.5$ & $p_{\perp}>0.8$ \\ \hline
case 1 & $K^+$ & 0.87 & 0.32 & 0.25 & 0.30 \\
       & $K^-$ & 0.68 & 0.19 & 0.24 & 0.25 \\ \hline
case 2 & $K^+$ & 0.72 & 0.17 & 0.21 & 0.34 \\
       & $K^-$ & 0.55 & 0.08 & 0.20 & 0.27 \\ \hline
\end{tabular}
\end{table}

%
%
\pagebreak

\begin{figure}
\caption{RQMD simulation of central Pb+Pb collisions
(impact parameter $b=0$ fm) at 158 GeV/nucleon: a high density domain
is created at an initial stage ($t=1$ fm in the center of mass frame)
by moving 10\% (case 1) or 20\% (case 2)
hadrons into an elliptical cylinder at the center
of the fireball. 
Particle density at $t=1$ fm projected on the $z=0$ plane is shown
at the left.
The high energy density domain is shown in the sketched ellipse.
Anisotropy in configuration space is shown at right for the two
initial conditions studied.}
\label{fig:conf}
\end{figure}

\begin{figure}
\caption{Average energy density in the cylinder (in a typical event)
as a function of time 
in the center of mass frame. The solid curves are energy density calculated
from all particles. The dashed curves are energy density calculated from
particles with rapidity $|y|<1.5$. The lower solid curve and the lower
dashed curve are the corresponding energy density from a default RQMD event.
The increase in the energy density at $t=1$ fm corresponds to creation of
the high energy density domain (case 2).
}
\label{fig:e_vs_t}
\end{figure}

\begin{figure}
\caption{Azimuthal distribution of charged pion multiplicity at freeze-out
for case 1 (top) and case 2 (bottom).
Only pions with rapidity $|y|<2$ are included. $\phi=0$ corresponds to the
major axis ($+x$ in Fig.~\ref{fig:conf}) 
of the ellipse of the high energy density domain.
The solid curve is a fit to the functional form~(\ref{eq}).
The dotted line is the average charged pion multiplicity in default RQMD
events.}
\label{fig:pion}
\end{figure}

\begin{figure}
\caption{Azimuthal distribution of charged pion multiplicity at freeze-out
in rapidity $|y|<2$ for various $p_{\perp}$ cuts for case 2.}
\label{fig:pion_pt}
\end{figure}

\begin{figure}
\caption{Azimuthal distribution of $K^+$ (left) and $K^-$ (right)
multiplicity at freeze-out in rapidity $|y|<2$
for various $p_{\perp}$ cuts for case 2.}
\label{fig:kaon}
\end{figure}

\begin{figure}
\caption{Azimuthal distribution of the difference between $K^+$ and $K^-$
multiplicities (top) and $\Lambda$ multiplicity (bottom) at freeze-out
in rapidity $|y|<2$ for case 2.}
\label{fig:lambda}
\end{figure}

%
%
\pagebreak

\centerline{\epsfxsize=6in\epsfbox[0 0 600 700]{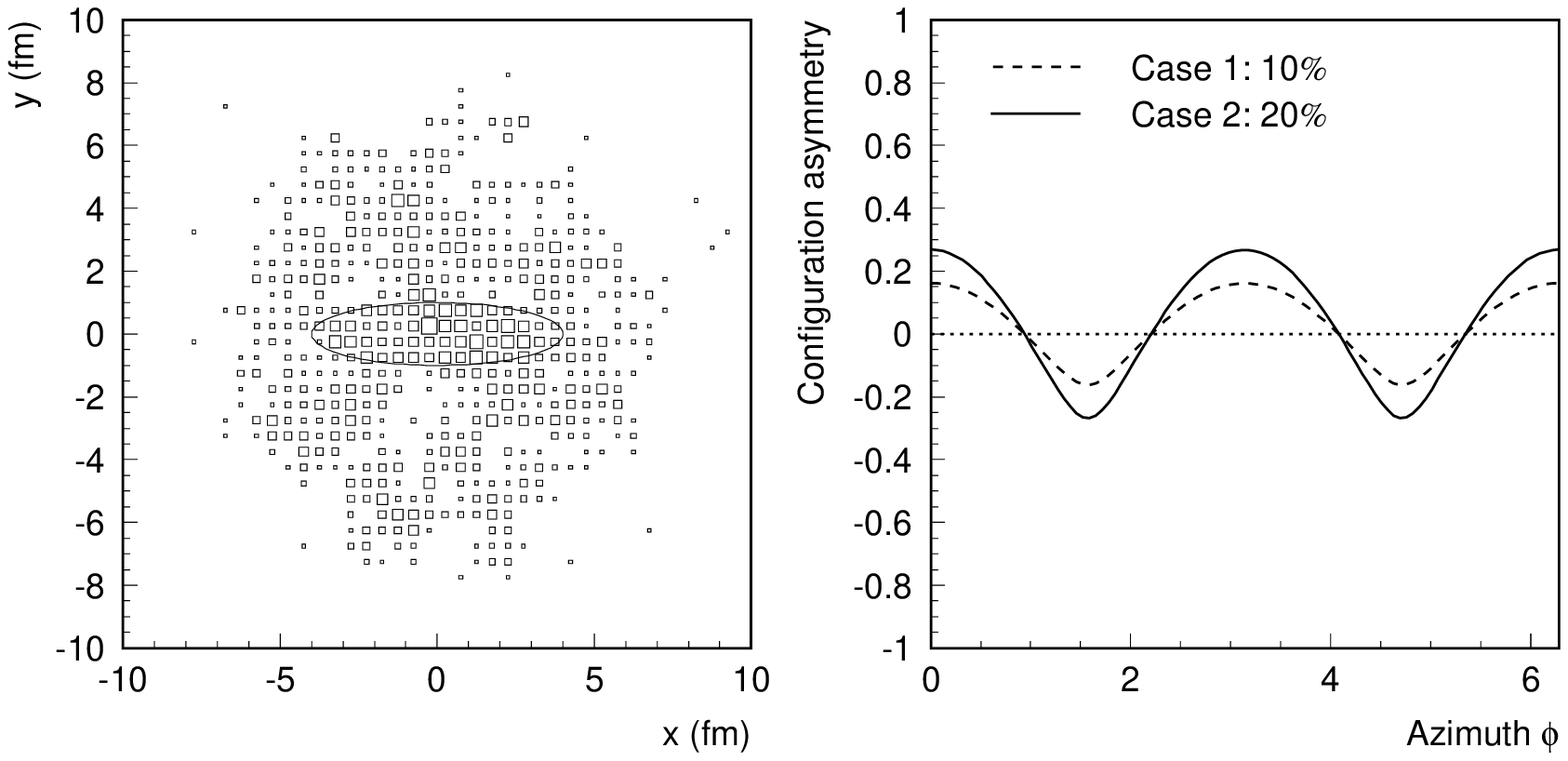}}
\bigskip
\centerline{Figure \ref{fig:conf}}
\pagebreak

\centerline{\epsfxsize=6in\epsfbox[0 0 600 700]{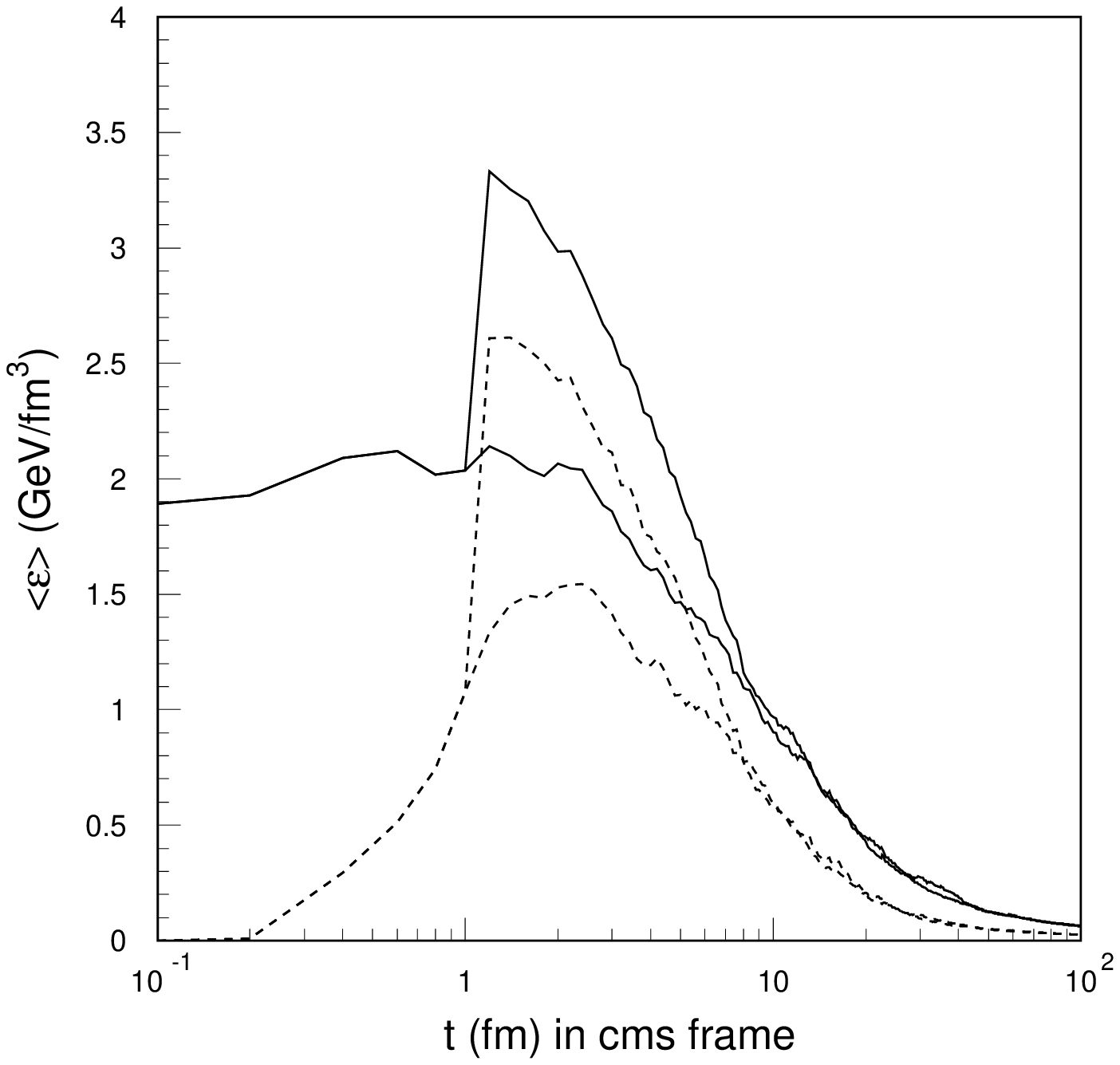}}
\bigskip
\centerline{Figure \ref{fig:e_vs_t}}
\pagebreak

\centerline{\epsfxsize=6in\epsfbox[0 0 600 700]{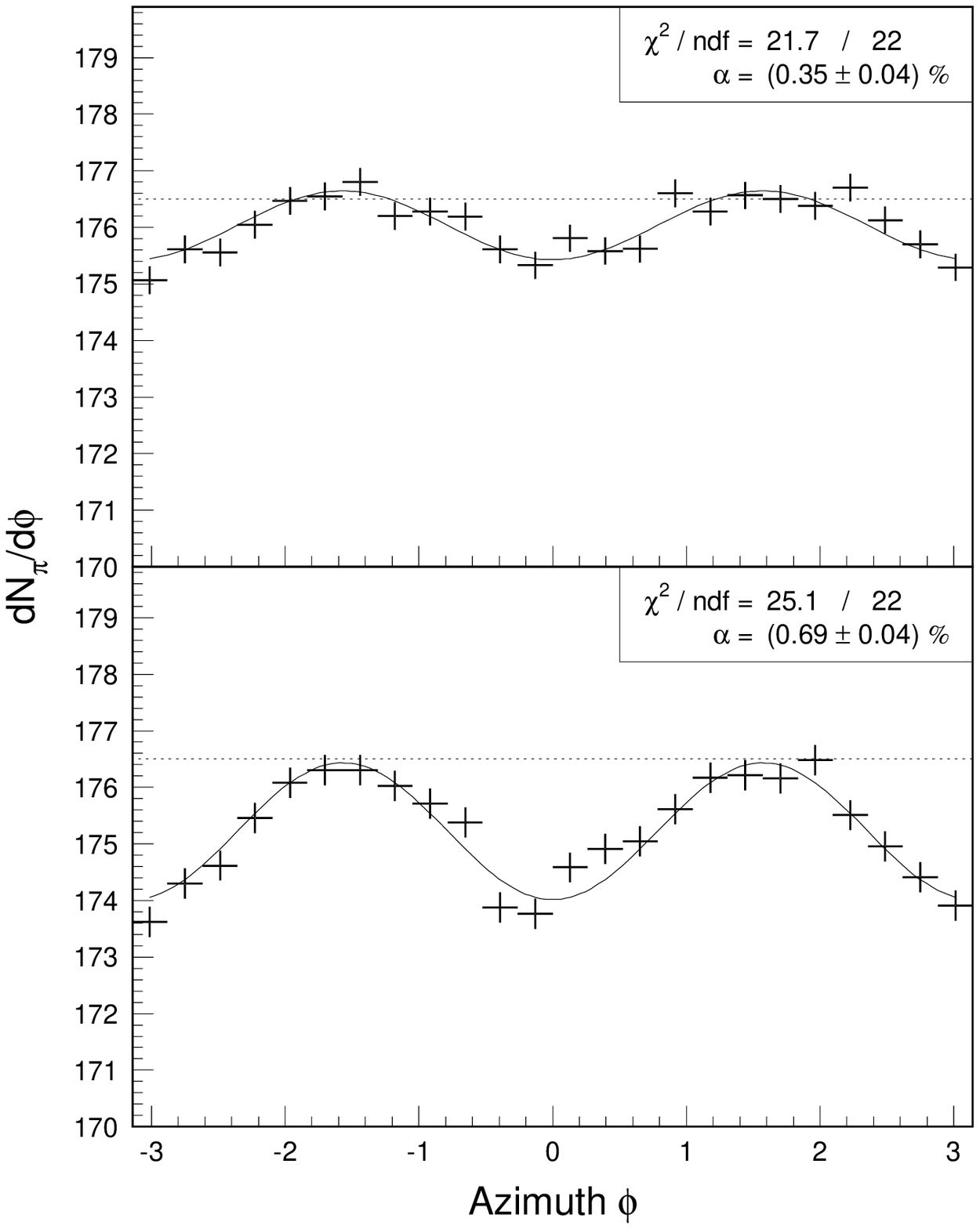}}
\bigskip
\centerline{Figure \ref{fig:pion}}
\pagebreak

\centerline{\epsfxsize=6in\epsfbox[0 0 600 700]{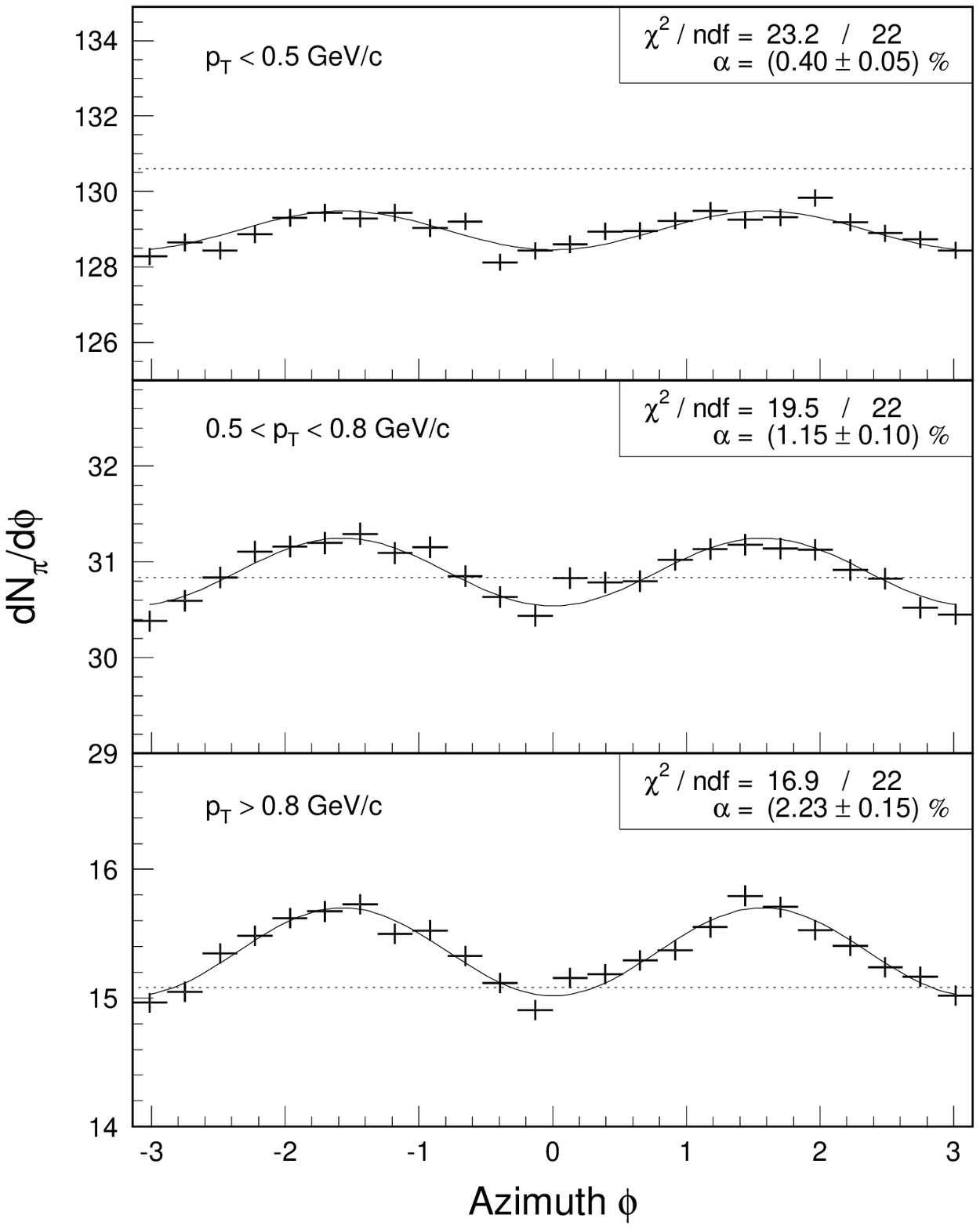}}
\bigskip
\centerline{Figure \ref{fig:pion_pt}}
\pagebreak

\centerline{\epsfxsize=6in\epsfbox[0 0 600 700]{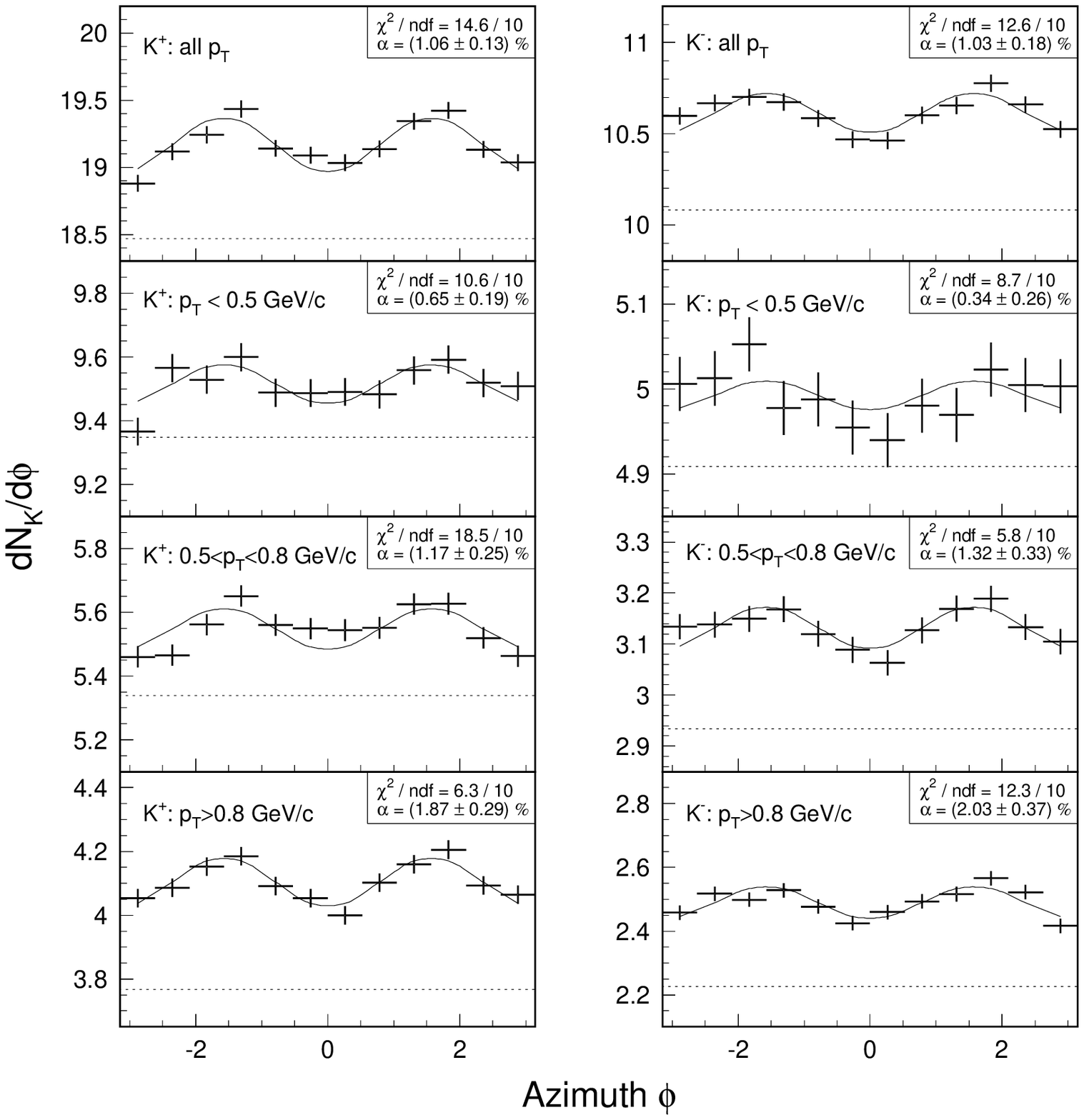}}
\bigskip
\centerline{Figure \ref{fig:kaon}}
\pagebreak

\centerline{\epsfxsize=6in\epsfbox[0 0 600 700]{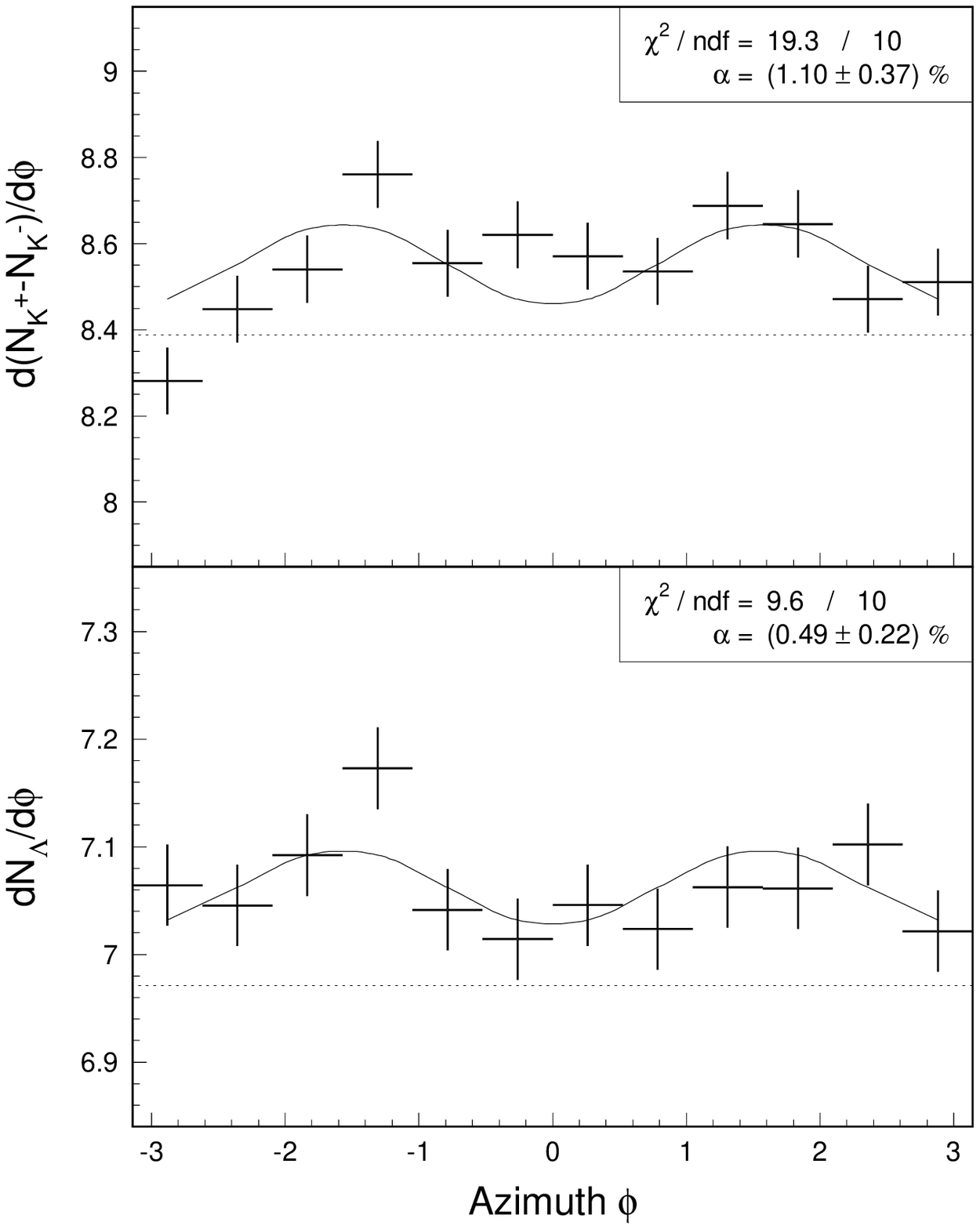}}
\bigskip
\centerline{Figure \ref{fig:lambda}}

\end{document}